\documentclass[journal,final,twocolumn]{IEEEtran}
\usepackage{amsmath,bm}
\usepackage{array}
\usepackage{graphicx}
\usepackage{xcolor}
\usepackage{amsfonts}
\usepackage{hyperref}
\usepackage{xcolor}
\usepackage{graphics}
\newcommand{\dcn}{{\textsc{DualCoreNet }}}
\newcommand{\dcnn}{{\textsc{DualCoreNet}}}

\usepackage{subfigure}
\usepackage{cite}

\usepackage{amsmath,amssymb,amsfonts}
\usepackage{algorithmic}
\usepackage{graphicx}
\usepackage{textcomp}
\usepackage{import}
\usepackage{subfigure}
\usepackage[super]{nth}
\usepackage{xcolor}
\usepackage{booktabs}
\usepackage{multirow}
\usepackage{pifont}
\newcommand{\cmark}{\ding{51}}
\newcommand{\xmark}{\ding{55}}
\usepackage{amsmath}

\begin{document}
\title{Dual Convolutional Neural Networks for Breast Mass Segmentation and Diagnosis in Mammography}
\author{
Heyi Li, Dongdong Chen, William H. Nailon,  Mike E. Davies \IEEEmembership{Fellow, IEEE}, and David Laurenson
\thanks{E-mail: \{heyi.li, d.chen\}@ed.ac.uk. H. Li, D. Chen, M. Davies and D. Laurenson are with the School of Engineering, University of Edinburgh, UK. W. Nailon is with Oncology Physics Department, Edinburgh Cancer Centre, Western General Hospital, Edinburgh, UK. D. Chen and M. Davies are funded by the ERC Advanced grant 694888, C-SENSE.}}
\maketitle

\begin{abstract}
Deep convolutional neural networks (CNNs)  have emerged as a new paradigm for Mammogram diagnosis.
Contemporary CNN-based computer-aided-diagnosis (CAD) for breast cancer directly extract latent features from input mammogram image and ignore the importance of morphological features.
{In this paper, we introduce a novel deep learning framework for mammogram image processing, which computes mass segmentation and simultaneously predict diagnosis results.}
Specifically, our method is constructed in a dual-path architecture that solves the mapping in a dual-problem manner, with an additional consideration of important shape and boundary knowledge. 
One path called the Locality Preserving Learner (LPL), is devoted to hierarchically extracting and exploiting intrinsic features of the input. Whereas the other path, called the Conditional Graph Learner (CGL) focuses on generating geometrical features via modeling pixel-wise image to mask correlations.
By integrating the two learners, both the semantics and structure are well preserved and the component learning paths in return complement each other, contributing an improvement to the mass segmentation and cancer classification problem at the same time. 
We evaluated our method on two most used public mammography datasets, DDSM and INbreast. Experimental results show that \dcn achieves the best mammography segmentation (in both high and low resolution) and classification simultaneously, outperforming recent state-of-the-art models. 
\end{abstract}

\begin{IEEEkeywords}
Mammography Diagnosis, Dual-Path Network, Deep Learning
\end{IEEEkeywords}

\section{Introduction}
\label{sec:introduction}
\begin{figure}[!t] 
    \centering
    \subfigure[Annotation of a benign mass]
    {\includegraphics[width=0.2\textwidth]{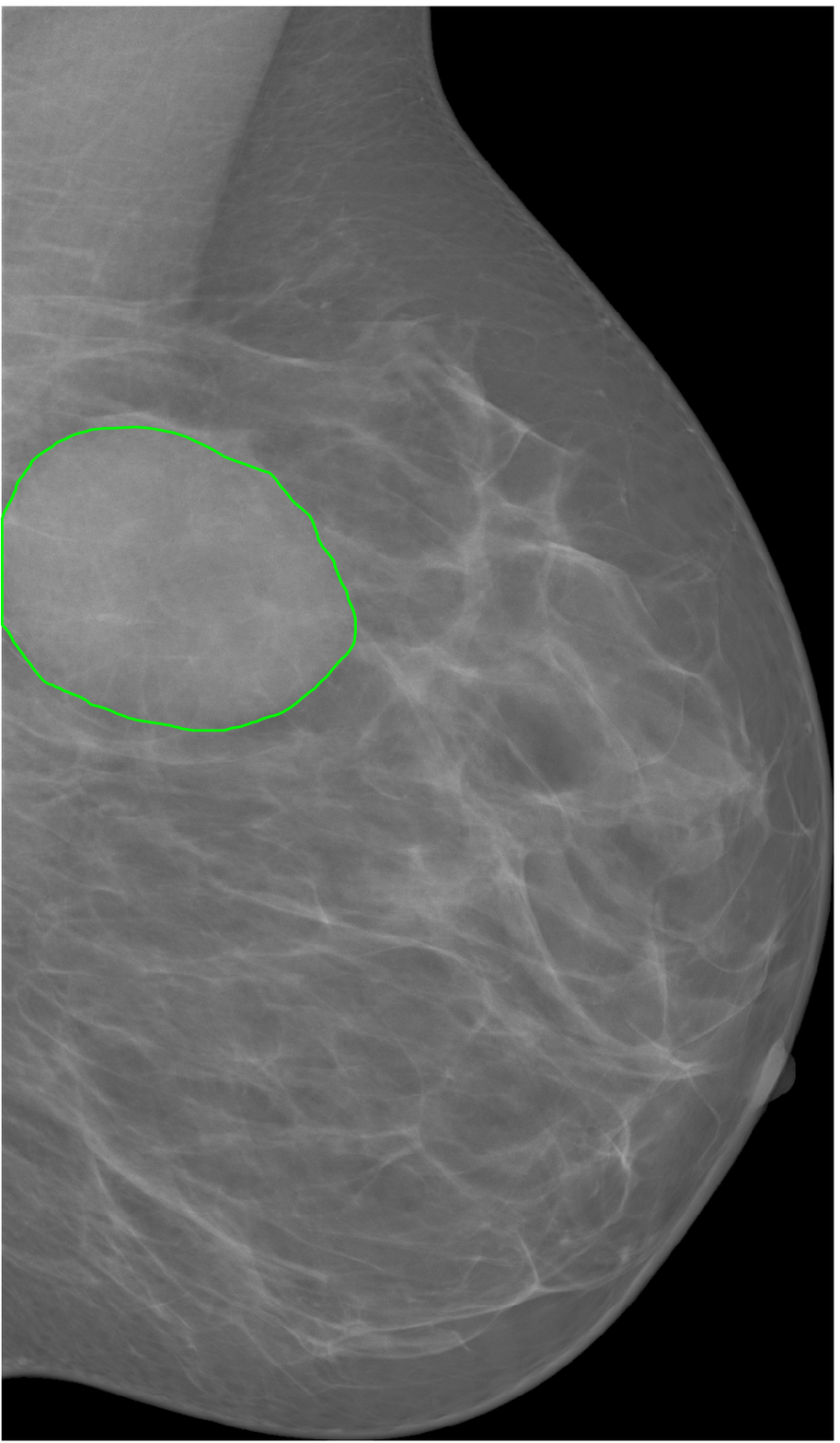}
        \label{fig:manul_inspection_benign}}
    \hfil
    \subfigure[Annotation of a malignancy]
    {\includegraphics[width=0.2\textwidth]{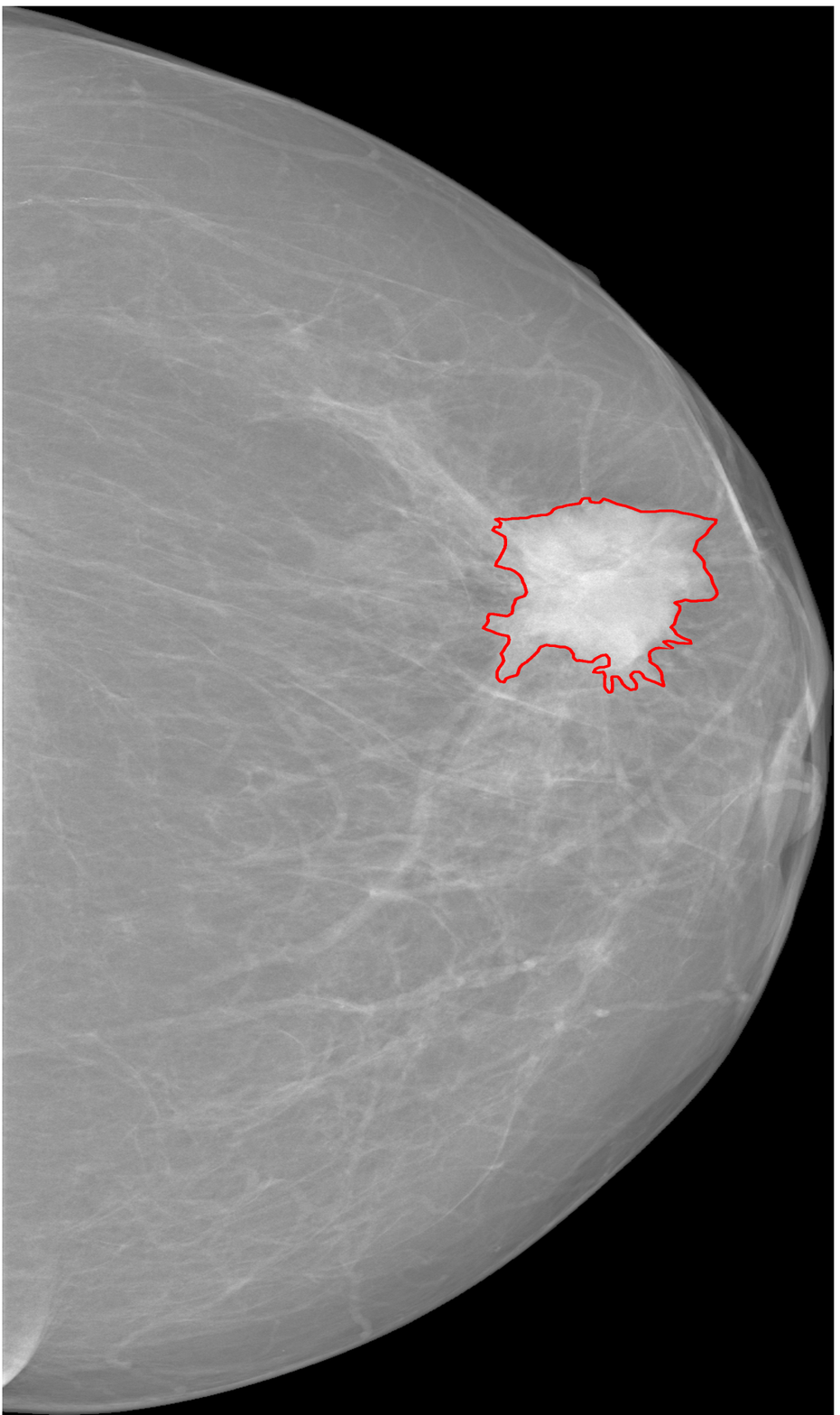}
        \label{fig:manul_inspection_malignant}}
  \caption{Manual inspection examples for breast masses in full field digital mammography (FFDM) INBreast dataset \cite{moreira2012inbreast}. 
    (a) contains a benign mass delineated with green lines, which is of oval shape and circumscribe boundaries in a CC view. 
    (b) shows a malignant mass in red lines captured in a MLO view, which is of irregular shape and spiculated boundaries.}
  \label{fig:manul_inspection} 
\end{figure}

\IEEEPARstart{B}{reast} cancer, according to the International Agency for Research on Cancer \cite{IARC2008}, is the most frequently diagnosed cancer.
Screening mammography is widely employed and has shown its significance especially for invasive breast tumours when they are too small to be palpable or cause symptoms. The manual inspection of a mammogram typically requires the lesion's identification as either benign or malignant, and sometimes the according delineation.  However, the manual inspection is tedious, subjective, and prone to errors \cite{oliver2010review, shen2019deep, swiderski2017novel}. 
Striving for the optimal health care, mammographical computer aided diagnosis (CAD) systems are designed as an alternative to a double reader, aiming to achieve similar inspection results to that of human experts (Fig. \ref{fig:manul_inspection}).

\begin{figure*}[!t]
    \centering
    \includegraphics[width=0.7\textwidth]{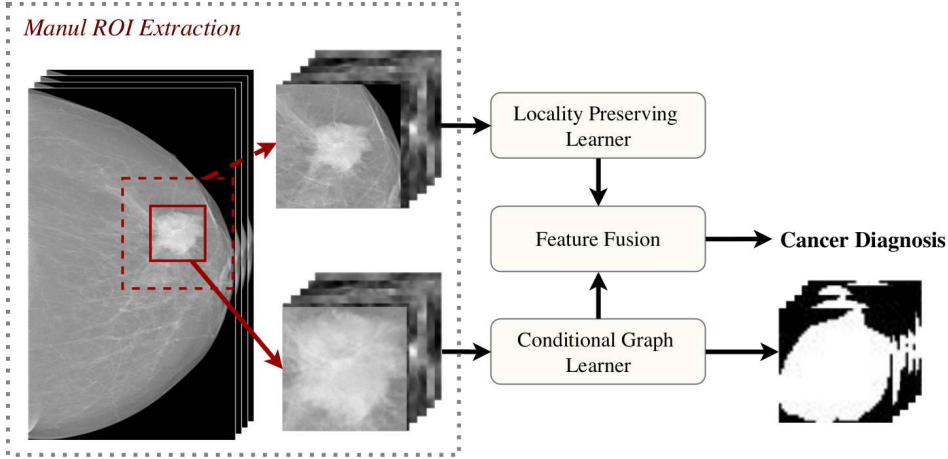}
    \caption[The flow diagram of \dcn]{The flow diagram of proposed \dcnn. With the extracted multi-scale ROIs as the input of LPL and CGL path separately, the \dcn outputs both segmentation mask and diagnosis label.}
    \label{fig:flowchart_dcn}
\end{figure*}

Many conventional machine learning algorithms have been proposed to tackle this  problem, which typically comprises various image processing operations (such as image segmentation, feature extraction, feature selection, and classification). 
The performance of a conventional CAD often relies heavily on cumbersome hand-engineered features \cite{varela2006use}, which are subsequently introduced into various classifiers. 
Oliver et al. \cite{oliver2010review} has demonstrated in their review paper that an accurate segmentation is the foundation of subsequent cancerous diagnosis, since the likelihood of malignancy depends on the shape and margin of lesions \cite{jalalian2013computer}. 
This statement has been empirically verified by a number of works \cite{domingues2012inbreast, varela2006use, dhahbi2015breast, swiderski2017novel}, which all claim that the most accurate breast mass diagnosis was obtained by the shape-related descriptors when compared with other conventional hand-crafted features. 
In fact,  traditional machine learning algorithms are still popular in recent commercial CADs. However, there is significant room for improvement, especially for breast mass diagnosis. 

In recent years, leveraging the insights from the success of deep neural networks (deep learning) \cite{lecun2015deep} in computer vision tasks \cite{chen2017graph,zhu2017deep,shams2018deep,chen2017unsupervised,golbabaee2018deep,chen2019deep,carneiro2017automated,chen2020compressive}, a noticeable shift to  deep learning based CADs has been seen. 
Some works have proposed the use of extracting segmentation-related features by CNNs with radiologists' pixel-level annotations, in order to further improve automatic diagnosis performance \cite{dhungel2016automated, dhungel2017deep}. 
However, this method requires large volume of accurate pixel-wise annotations, which are very difficult to obtain in practice. 
In order to enhance the network without using binary mask labeling, some authors explored the  performance with an automatic segmentation algorithm \cite{dhungel2017deep}. Yet this automatic setup has caused a considerable performance drop.
The poor performance is likely caused by the multi-stage process training.
Based on these observations we are motivated to construct a CNN architecture trained in an end-to-end fashion, in order to \emph{jointly} solve the breast cancer diagnosis (benign vs malignant) \emph{and} the segmentation problem in mammography.

In this paper, we presented a multi-scale dual-path CNNs as shown in Fig. \ref{fig:flowchart_dcn}, 
to solve the image to diagnosis label mapping in a dual-problem manner. 
In particular, 
the dual-problem here especially refers to the segmentation and classification problem. A preliminary version of this work appeared in \cite{li2019deep}. This paper extends \cite{li2019deep} by providing a more detailed description of the work and more comprehensive experimental evaluations.  Based on the accurate breast mass segmentation algorithm presented in \cite{li2018improved} and the related breast mass classifiers \cite{carneiro2017automated, arevalo2016representation, kooi2017discriminating, dhungel2016automated, li2019signed}, a \textbf{Dual}-path \textbf{Co}nditional \textbf{Re}sidual \textbf{Net}work (\dcnn) for mammography analysis is introduced. 
Firstly, a mass and its context texture learner called the \textit{Locality Preserving Learner (LPL)} is built with stacks of convolutional blocks, achieving a mapping from relative large scale ROIs to class labels. 
Secondly, an integrated graphical and CNN model, called the \textit{Conditional Graph Learner (CGL)} is employed to learn the relative small scale ROI  to mask correlation, and the extracted segmentation features will be further used to improve the final mass classification performance. 
Additionally, we train the model with multi-scale ROIs, since the surrounding tissues of breast masses play a pivotal role for the accurate cancer diagnosis, whereas the contextual tissues are less irrelevant for the segmentation task.
For a certain breast mass, a larger scale ROI is used as the input of LPL for richer features extraction, and a smaller scale ROI is employed by the CGL path. 
\dcn achieves the best mammography segmentation and classification simultaneously, outperforming recent state-of-the-art models. 
The main contributions of this paper are the following: 
\begin{enumerate}
    \item To our knowledge, \dcn is the first fully automatic dual-path CNN-based mammogram analysis model that takes advantage of an automatic segmented mask for mass classification;
	\item Our method has achieved the best performance for breast mass segmentation in both low and high resolutions;  
	\item \dcn has achieved comparable results with mass classification tasks on publicly available mammography datasets.
\end{enumerate}

\textbf{Organization}. The rest of the paper is arranged as follows: Section \ref{sec:2} presents the related preliminary techniques,  Section \ref{sec:3} introduces the proposed \dcn methodology, and Section \ref{sec:4} shows the experimental results. We conclude this paper in Section \ref{sec:5}. 


\section{Preliminary}
\label{sec:2}
In this section, we will introduce three  machine learning methods, which are related to our proposed \dcnn. 
Firstly, we will discuss the residual learning and the inception modules separately, which have improved the deep model generalization in CNNs.  Then the conditional random Field (CRF), a type of graphical model, will be discussed for the medical image segmentation. 

{\begin{figure*}[!t]
    \centering
    \includegraphics[width=1\textwidth]{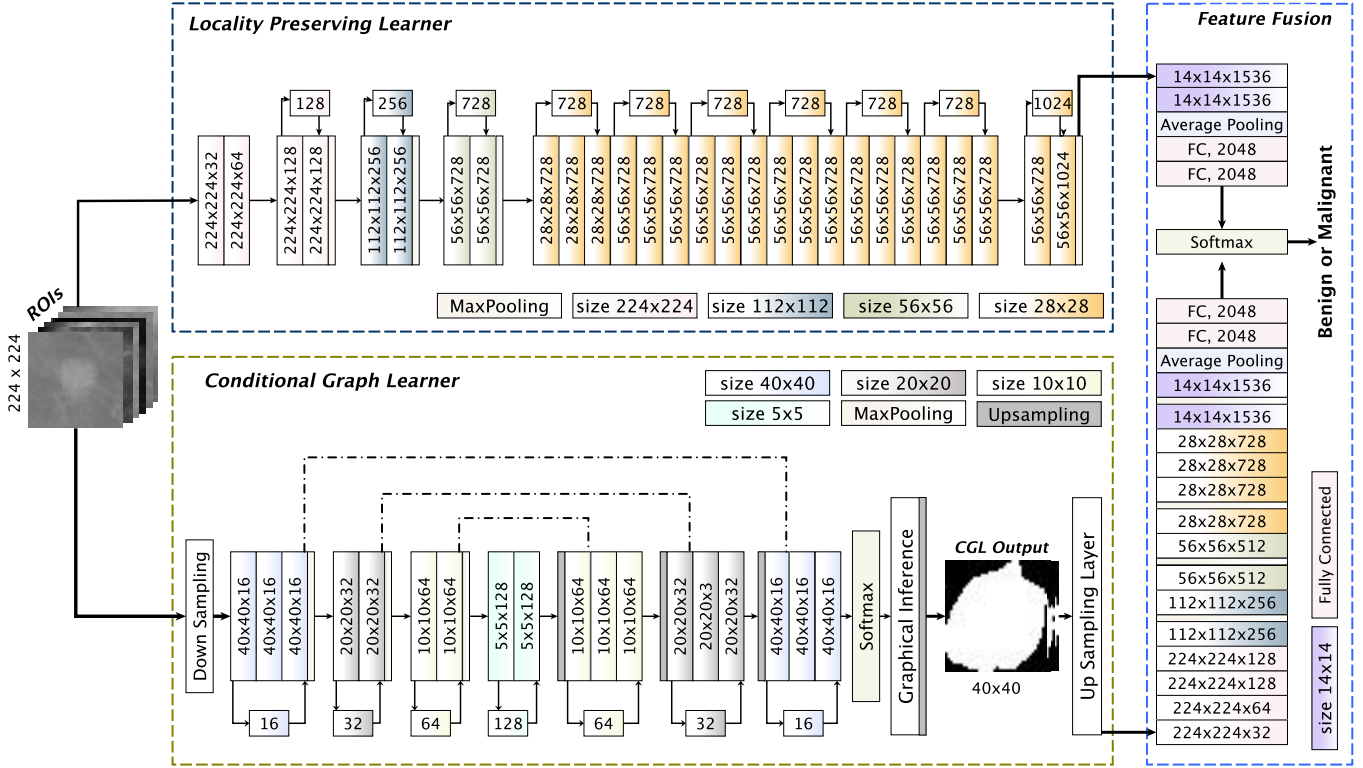}
    \caption{Overall architecture of our proposed \dcn architecture.}
    \label{fig:dcn_architecture}
\end{figure*}}

\subsection{Residual Learning}
Residual learning \cite{he2016deep} is proofed a efficient way to accelerate the neural network training and avoid the gradient vanishing/exploding problems, which had been extensively applied method for myriad computer vision tasks. The main idea of of residual learning is the use of residual connections between the input and output to the neural network internal layers or blocks or even entire network. 
In particular, by letting the desired input to output mapping in a residual module as $\mathcal{H}(\mathbf{x})$, the residual function obtained in each module is defined with:
\begin{equation}
\mathcal{H}(\mathbf{x}):=\mathbf{x} + \mathcal{F}(\mathbf{x}).
\end{equation} 
The gradients of a residual module in each layer are therefore pre-conditioned to be close to the identity function, solving the gradients vanishing problem. 
In this way, CNNs can be constructed with many more layers with efficacious training.  
Specifically, residual connections enable the deep neural network to learn the residual between the input and output of a module, instead of directly learning the mapping.


{\begin{figure*}
    \centering
    
    \subfigure[Module A]
    {
        \includegraphics[width=0.37\textwidth]{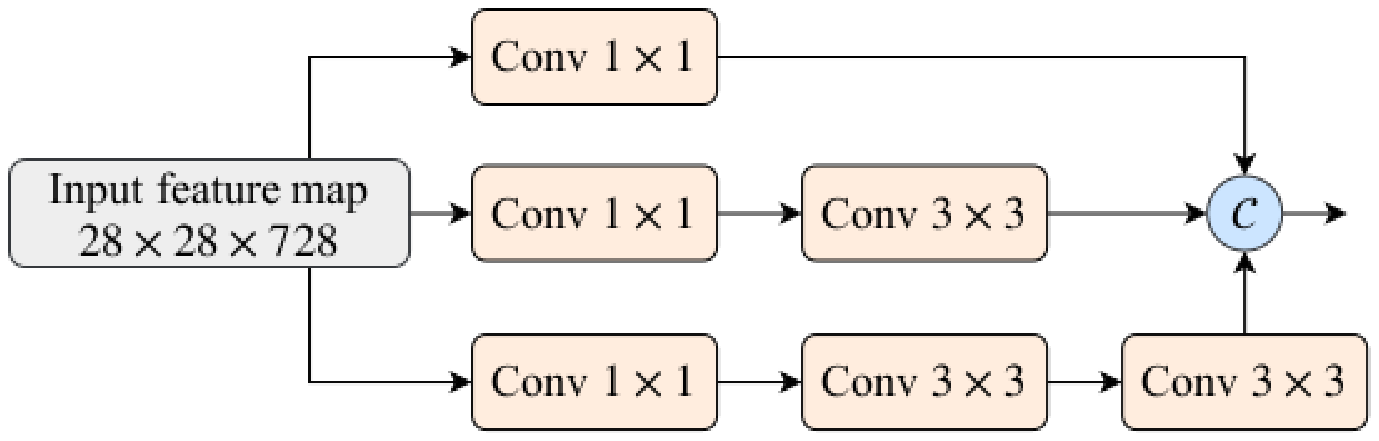}
        \label{fig:sconv_a}
    } 
    \subfigure[Module B]
    {
        \includegraphics[width=0.37\textwidth]{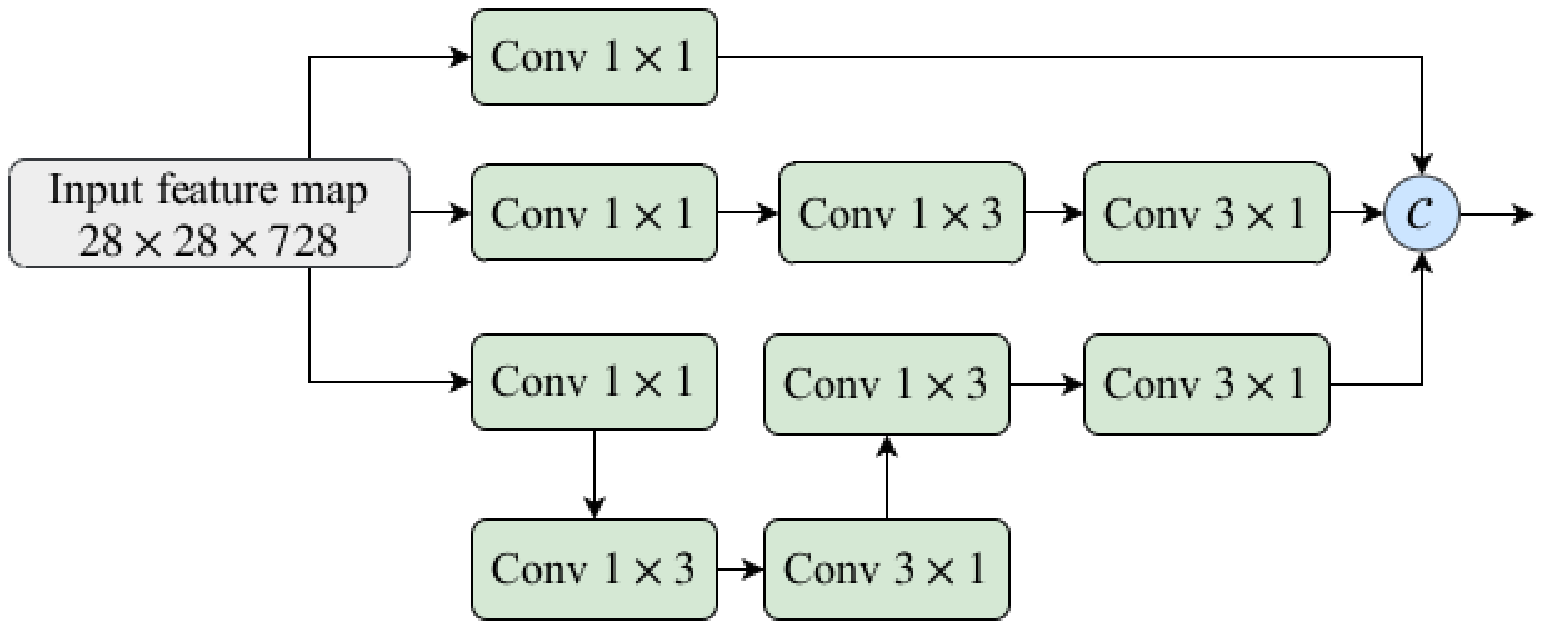}
        \label{fig:sconv_b}
    }\\
    \subfigure[Module C]
    {
        \includegraphics[width=0.37\textwidth]{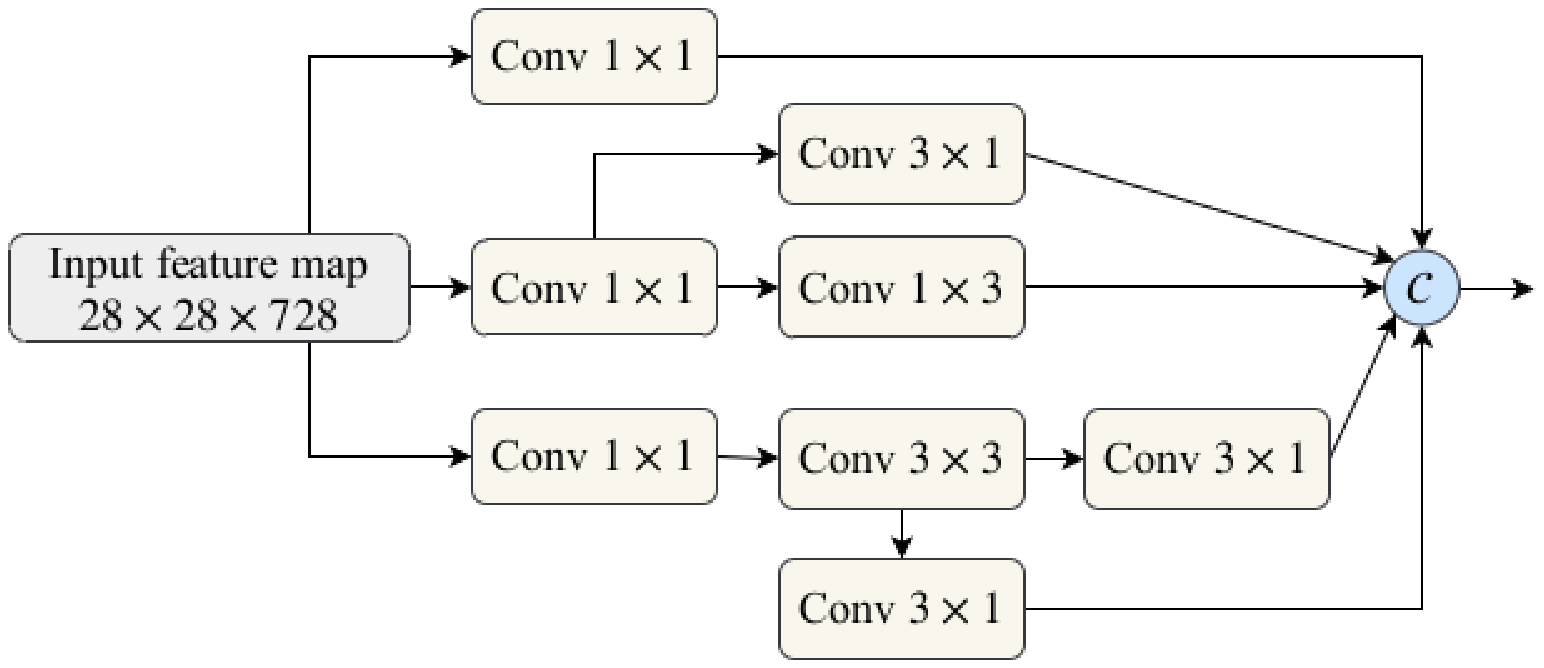}
        \label{fig:sconv_c}
    }
    \subfigure[Module D]
    {
        \includegraphics[width=0.37\textwidth]{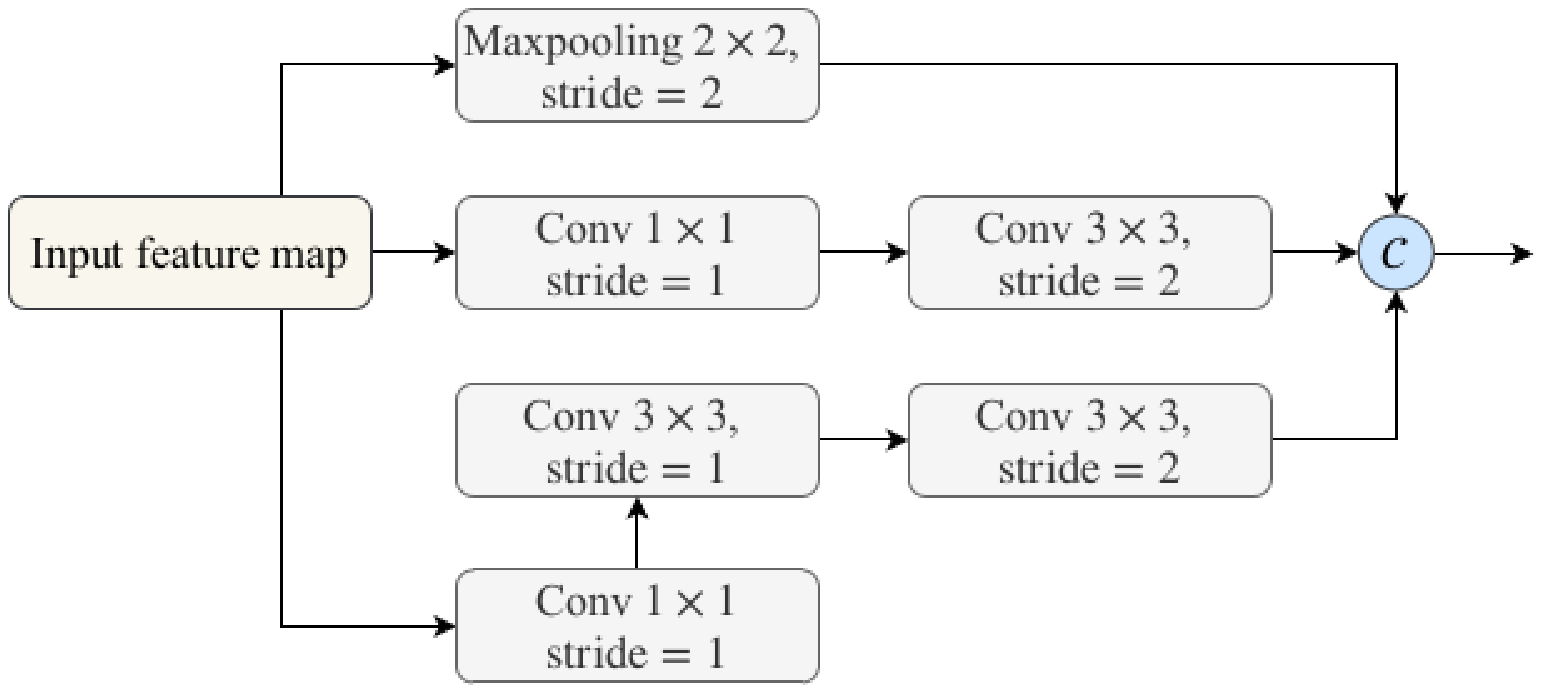}
    \label{fig:sconv_reduction}
    }
    \caption{Three convolution modules applied in \dcnn, where the original input feature map are all of the dimension $28\times28\times728$ and the expanded features are concatenated in each module. 
    (a) \emph{Module A} replaces a $5\times5$ convolution with two $3\times3$ convolutions.
    (c)\emph{Module B} replaces a $7\times7$ with smaller kernel convolutions. (c) \emph{Module C} expands a filter bank of $8 \times 8$ grids.
    (a) \emph{Module D} is a dimension reduced convolution module, which halves the module input's spatial dimension.}
    \label{fig:sconvs}
\end{figure*}}

\subsection{Conditional Random Fields}
Conditional Random Fields (CRF) \cite{krahenbuhl2011efficient}, as a variant of Markov Random Fields (MRFs),  incorporates the label consistency with similar pixels and provides sharp boundary and fine-grained segmentation.
Typically a CRF formulates the label assignment problem as a probabilistic inference problem, in which pixel labels are modeled as MRF random variables and conditioned upon the observations \cite{krahenbuhl2011efficient}.  
Given an observed deep latent feature $h(\mathbf{x})$, the Gibbs distribution of a fully connected CRF with $\mathcal{V}$ nodes and $\mathcal{E}$ edges is defined as:
\begin{equation}
    p(\mathbf{y}|h(\mathbf{x})) = \dfrac{1}{Z(h(\mathbf{x}))}\exp{\big(-\sum_{c\in\mathcal{C}}
    \phi_c(h(\mathbf{x}),{y}_c)\big)},
\end{equation}
where $\mathbf{y}$ is pixel-level label state,  $c$ is the one item in the clique set $\mathcal{C}$, $y_c$ is the clique joint label, $\phi_c$ is the feature vector, and $Z$ is the partition function.

Recently, due to its flexibility and efficiency, CRF has been applied on medical images for various segmentation problems \cite{zhu2018adversarial, kamnitsas2017efficient} in which the CRF was mainly used as a post processing step which usually lead to an inherent shrinking bias problem \cite{guo2020improving}. 
In contrast, in this paper, the CRF is used as a specific neural network layer that comprised of the unary potentials and the pair-wise potentials connecting all other pixels with the aim of modeling long-range connections in arbitrarily large neighbourhoods and simultaneously preserve the advantageous fast inference \cite{krahenbuhl2011efficient, kamnitsas2017efficient}, in order to provide a balanced partitioning.

\section{Methodology}
\label{sec:3}
In this section,  we will first define the notations used throughout the paper and formulate the problem to solve. After that, the two paths are separately discussed, which is followed by an introduction of two different feature aggregation methods.

\subsection{Notations and Problem Formulation}
Given $N$ mammograms and the biopsy-confirmed annotations from human experts,  the dataset can be then noted as $\{X, Y, \mathcal{Z}\} = \{\{\mathbf{x}^{(n)}\}, \{\mathbf{y}^{(n)}\}, \{z^{(n)}\}\}_{n=1}^N$, where 
$\mathbf{x}^{(n)}  \in \mathbb{R}^{H\times W}$ represents the $n^{\textrm{th}}$ mass-contained ROI with spatial dimension $H\times W$, $\mathbf{y}^{(n)}  \in \{0:\textrm{Normal pixel},\ 1:\textrm{Mass pixel}\}^{H\times W}$ is the pixel-level annotation corresponding to the $n^{\textrm{th}}$ ROI, and $z^{(n)}\in\{0, 1\}$ is a scalar indicating the diagnosis class label, and ``0'' and ``1''  represents the benign and the malignant class, respectively. Note that the cropped ROIs contain only one mass in our scenario.
The main targets solved by \dcn can be formulated as follows:
(1) given a mass contained mammogram ROI, \dcn is desired to map the original images to binary masks, so that mass pixels are segmented: $\mathbf{x}^{(n)} \rightarrow \mathbf{y}^{(n)}$.  
(2) With the original mammogram images and the obtained pixel-level labels, learn a nonlinear mapping to the diagnosis label: ${\mathbf{x}^{(n)}, \mathbf{y}^{(n)}} \rightarrow \mathbf{z}^{(n)}$, where
.

\subsection{Motivation}
Practically, radiologists make decisions about breast masses with their shape and boundary features. The more irregular the shape is, the more likely the mass is malignant \cite{dhungel2015deep}, i.e., the classification results generally heavily rely on the segmentation results \cite{oliver2010review}.
In analogy, decoupling a complicated learning task into several sub-tasks that are easier to solve is has also been proved an efficient learning paradigm in machine learning, there many methods aim to use multi-paths neural networks to solving image classification or other image restoration problems \cite{liu2020improved,wang2019camp,chen2019deep,zheng2020dual,fan2019birnet,kim2019multi,gao2017classification,liu2018multi,ciompi2017towards,Liang2019multi}. However, the decoupling of breast mass diagnosis problem is seldom studied and the multi-path architecture has never been exploited with the dual segmentation and classification problem. Based on these, we aim to close this gap by solving these two problems in one, thus further improving the mammography analysis.

\subsection{The proposed \dcn}

In this paper, we propose the \dcn architecture, which decouples the  differentiation of benign and malignant classes into dual problems:  segmentation and  classification. 
In the classification task, each input ROI sample (with surrounding tissues) will be classified into cancer category or not; whereas in the segmentation task, each pixel is  labeled as either $0$ or $1$ so that mass pixels can be accurately identified within the tight bounding box ROI. 

The \dcn takes a batch of multi-scaled mammogram ROIs as the input and outputs the mass segmentation masks and the diagnosis labels simultaneously (Fig. \ref{fig:flowchart_dcn}). 
The mass segmentation computes the mapping from smaller scale ROI to binary masks, i.e. $X \mapsto Y$. 
The mass classification solves the mapping of $X \mapsto Z$, by which the larger scale ROIs are mapped into  diagnosis labels.  
Based on this idea, the \dcn is constructed, which is comprised of the \emph{Locality Preserving Learner} (LPL) and the \emph{Conditional graph learner} (CGL) paths.

\begin{figure*}[!t]
    \centering
    \subfigure[The validation loss of DDSM]
    {
        \includegraphics[width=0.45\textwidth]{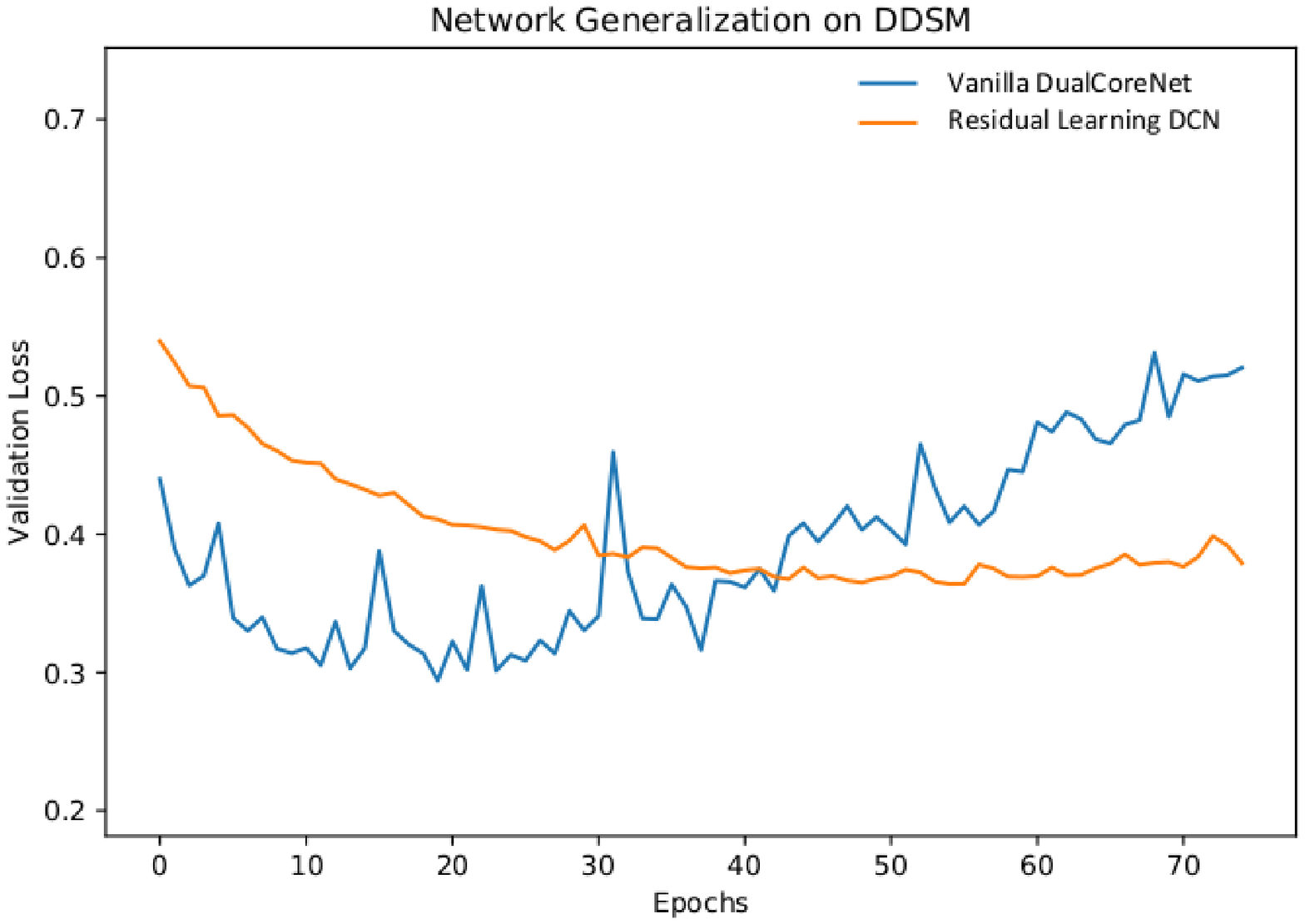}
    } 
    \subfigure[The validation loss of INbreast]
    {
        \includegraphics[width=0.475\textwidth]{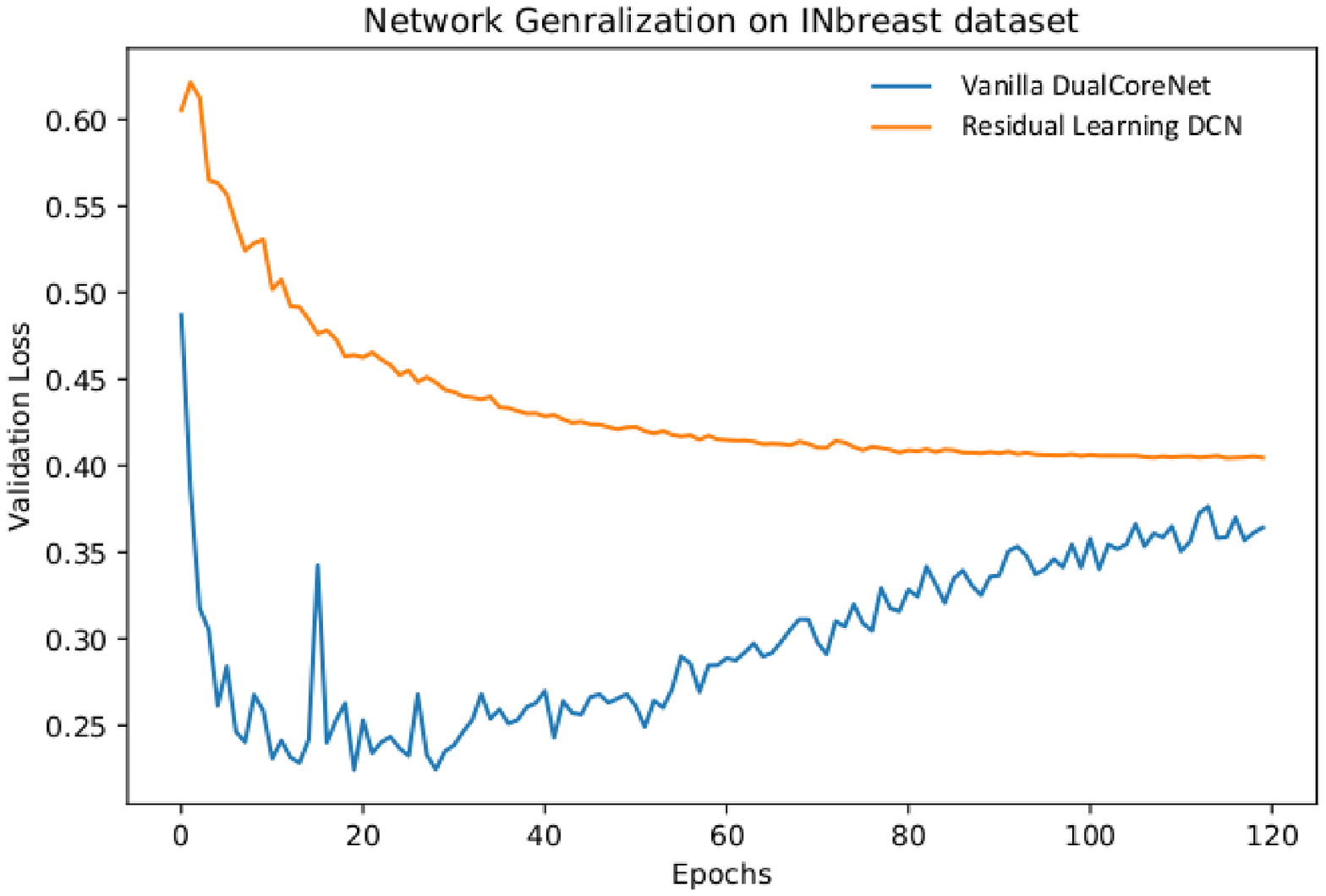}
    }
   \caption{The residual learning and vanilla \dcn validation loss on DDSM and INbreast datasets.}
    \label{fig:validation_loss}
\end{figure*}
\begingroup
\begin{table*}[!t]	\renewcommand{\arraystretch}{1.2}
	\caption[High-resolutional breast Cancer local analysis performance comparison]{High-resolutional breast cancer local analysis with both segmentation and classification performance of state-of-art algorithms. The segmentation performance is assessed by the DI (\%) matrix and the classification performance (Malignant vs Benign) is evaluated with the AUC score. The ``Processed'' column represents the pre-processing and post-processing of the list algorithms respectively.}
	\centering
	\renewcommand{\arraystretch}{1.1} 
	\begin{tabular}{l|c|c|c|c}
		\toprule
		\mdseries Methodology & \mdseries Dataset & \mdseries Segmentation & \mdseries Classification & \mdseries End-to-end \\
		\midrule
		Arevalo \textit{et al.} \cite{arevalo2016representation} (2016) & private & - & 0.82 & \xmark \\
		\midrule
		Dhungel \textit{et al.} \cite{dhungel2016automated} (2016) & INbreast & - & 0.91 & \xmark \\
		\midrule
		Kooi \textit{et al.} \cite{kooi2017discriminating} (2017) & private & - & 0.80 & \xmark \\
		\midrule
		Dhungel \textit{et al.} \cite{dhungel2017deep} (2017) & INbreast & $85.0$ & 0.76 &  \xmark\\ 
		\midrule
		Al-antari \textit{et al.} \cite{al-antari2020deep} (2020) & INbreast & 92.36 & \textbf{0.95} & \xmark \\
		\midrule
		\multirow{2}{*}{\dcn}  & INbreast & \textbf{93.69} & 0.93  &\cmark \\
          & DDSM & \textbf{92.17} & \textbf{0.85} & \cmark\\
		\bottomrule
	\end{tabular}
	\label{tab:dcn_all_comparison}
\end{table*} 
\endgroup

\subsubsection{Locality Preserving Learner}

The LPL path is constructed to learn the hierarchical and local intrinsic features from large scale ROIs. Large scale ROIs includes both textural and contextual information which are pivotal for mass classification\cite{varela2006use, shen2019deep}. Inpisred by the well-known CNN backbone architecture, \emph{e.g.,} VGG \cite{VGG16and19_2014}, the ResNet \cite{he2016deep_ResNet1, he2016identity_ResNet2}, and the DenseNet \cite{huang2017densely}, in this paper we propose an effective architecture, especially for mammography diagnosis, as illustrated in Fig. \ref{fig:dcn_architecture}, and the several separable convolution modules Fig. \ref{fig:sconvs} along residual connections are employed. 

In this paper, the LPL path is constructed with 11 convolutional layers. In particular, the first four consecutive layers of the LPL employ naive convolutional layers along with the dimension reduction block (Fig. \ref{fig:sconv_reduction}), allowing the network to learn features at every chosen spatial scale. 
The number of feature maps  consistently increase in the first four convolutional layers, from 16  (input) to 728 (the \nth{4} layer output) feature maps, whereas the spatial dimension reduces from $224\times224$ to $56\times56$. 
Regarding the spatial downsampling, the maxpooling layer is employed in the \nth{2} layer, whereas the separable convolution in Figure \ref{fig:sconv_reduction} is utilized in the \nth{3} and \nth{4} layers.
Instead of using maxpooling followed by the convolutions, this  dimension reduction method concatenates different scaled features generated by one or two convolutional operators directly. 
After that,  two blocks of each \emph{Block-A} (\nth{5} and \nth{6} layers), \emph{Block-B} (\nth{7} and \nth{8} layers), and \emph{Block-C} (\nth{9} and \nth{10} layers) are separately constructed. 
These depth-wise separable convolutional layers produce the same number of feature maps and the identical spatial dimension, i.e. $28\times28\times728$.
By utilizing  $1 \times 1$ convolutions in the depth-wise separable blocks, the cross-channel correlations are learned first, resulting in a much smaller feature space.
Thereby, the LPL is enabled to learn richer features with much fewer parameters,  hence alleviating the overfitting problem markedly with the same amount of training data.

Lastly, the generated deep features in the \nth{11} layer  are activated by the softmax non-linearity.
The loss associated to the LPL layer is defined with categorical cross-entropy as:
\begin{equation}\label{eq:lpl}
    {{\text{\large\ensuremath\ell}}}_{\text{LPL}} = -\sum_{n=1}^N\log p({z}^{(n)} \mid \boldsymbol{x}^{(n)};\boldsymbol{\theta}_{1}\big)
\end{equation}
where $z$ is the class indicator and $\boldsymbol{\theta}_{1}$ is the corresponding parameter set in LPL. 

\begingroup
\begin{table}[!t]
\centering
\caption[Breast cancer diagnosis performance (Malignant vs Benign) of the LPL path in \dcn]{Breast cancer diagnosis performance (Malignant vs Benign) of the LPL path in \dcnn. }
\renewcommand{\arraystretch}{1.2}
\begin{tabular}{@{}cclll@{}}
\toprule
{Dataset} & {Pre-training} & {Augmentation}  & {AUC score}    \\
\midrule
DDSM & none & none & 0.73  \\
DDSM & none & flips & 0.73  \\
DDSM & none & flips, random crops & 0.74  \\
DDSM & ImageNet & none & 0.79  \\
DDSM & ImageNet & flips & 0.79  \\
DDSM & ImageNet & flips, random crops & \textbf{0.85} \\
\midrule
INbreast & none & none & 0.80  \\
INbreast & none & flips & 0.85  \\
INbreast & none & flips, random crops & 0.84  \\
INbreast & DDSM & none & 0.86  \\
INbreast & DDSM & flips & 0.89  \\
INbreast & DDSM & flips, random crops & \textbf{0.93}  \\
\bottomrule
\end{tabular}
\label{tab:dcn_lpl_classification}
\end{table}
\endgroup

\subsubsection{Conditional Graph Learner}

The CGL path aims to extract segmentation-related or geometrical features from the resulted binary mask produced by an image to pixel-level label  mapping.
However, adapting CNNs to pixel-level labelling tasks is a significant challenge, since convolutional filters produce coarse outputs and max-pooling layers further reduce the sharpness of segmented boundaries. 
Although many methods have been utilized for this problem \cite{ronneberger2015u, zheng2015conditional, li2018improved}, unfortunately the balanced partitioning with high pixel resolution is still a challenge to solve \cite{guo2020improving}. 

Thereby, we propose a novel breast mass segmentation CNN architecture for the CGL path, as shown in Fig. \ref{fig:dcn_architecture}, which is expected to not only precisely segment high-resolutional breast mass but also to control the model complexity. 
To do that,  a CRF inference layer is applied in the low resolutional latent space and a concatenation is added to connect the high-resolutional feature maps, so that exhaustive textural features are interlaced and fully used.

\begin{figure*}[!t]
    \centering
    \subfigure[DDSM]
    {\includegraphics[width=0.45\textwidth]{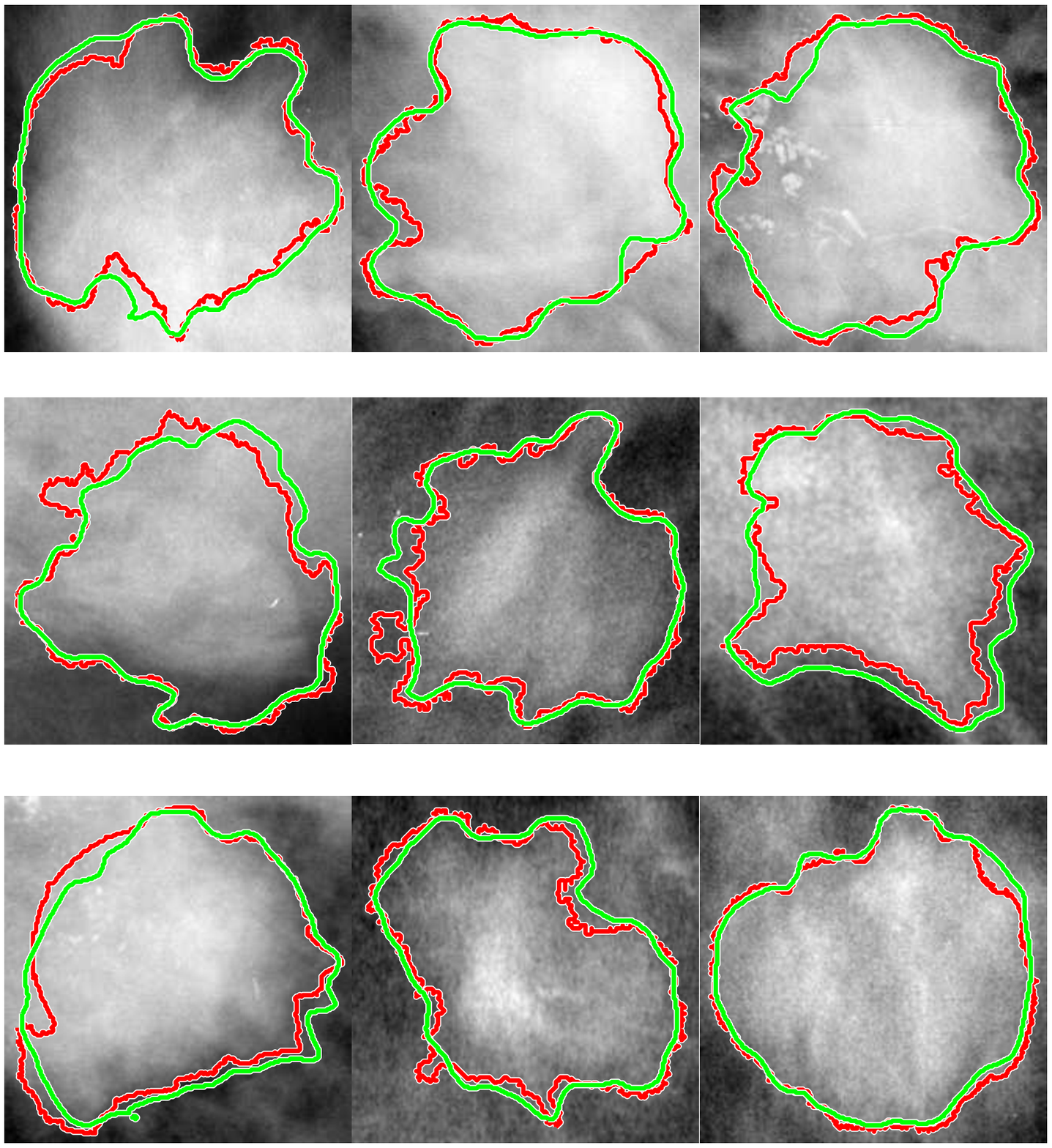}
        \label{fig:ddsm_contours}}
    \subfigure[INbreast]
    {\includegraphics[width=0.45\textwidth]{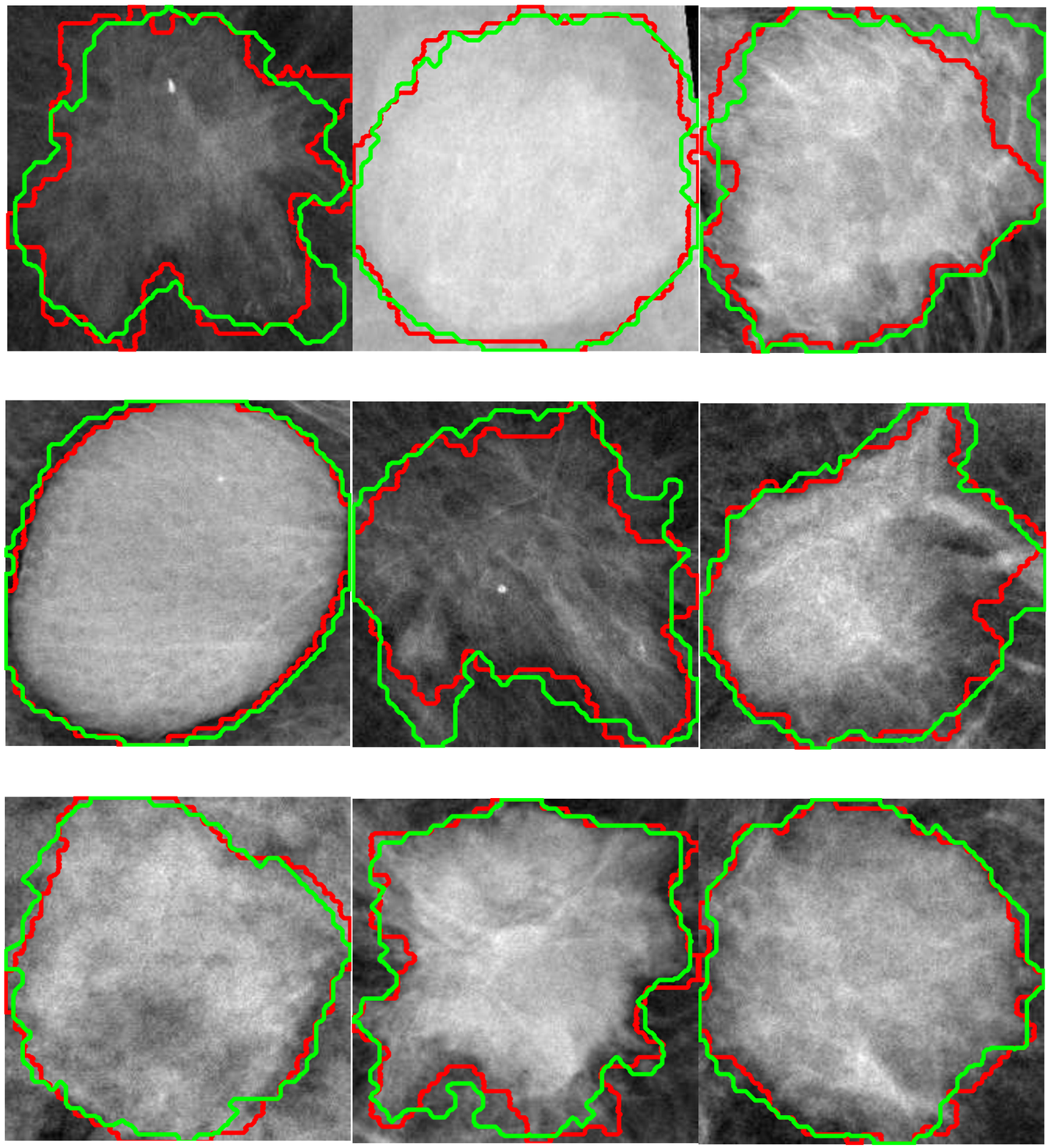}
        \label{fig:inbreast_contours}}
\caption{Exampled high-resolutional breast mass segmentation results on DDSM and INbreast datasets, with an visualized comparison between radiologists' annotation (red lines) and \dcn segmentation results (green lines).}
\label{fig:ddsm_contours}
\end{figure*}

{\begin{table*}[!t]	\renewcommand{\arraystretch}{1}
	\caption[Breast Mass Segmentation performance of state-of-art algorithms]{Quantitative breast mass segmentation performance (Dice coefficient, \%) of \dcn and several state-of-the-art methods on test sets. The ``Processed'' column represents for the pre-processing and post-processing of the list algorithms respectively.}
	\centering
	\renewcommand{\arraystretch}{1.2} 
	\begin{tabular}{c|c|c|c|c}
		\toprule
		\bfseries Methodology & \mdseries INbreast & \mdseries DDSM & \mdseries Spatial Dimension & \mdseries Pre/Post-processed \\ 
		\midrule 
		Beller {et al.} \cite{beller2005example} (2005) & - & $70$ & - & - / -\\ 
		Cardoso {et al.} \cite{cardoso2015closed} (2015) & $88$ & - & $40\times40$ & - / -\\ 
		Dhungel {et al.} \cite{dhungel2015deep1} (2015)& $88$ & $87$ & $40\times40$ & \cmark / \cmark\\ 
		Dhungel {et al.} \cite{dhungel2015deep} (2015)& $90$ & $90$ & $40\times40$ & \cmark / \cmark\\ 
		Zhu {et al.} \cite{zhu2018adversarial} (2018) & $89.36\pm0.37$ & $90.62\pm0.16$ & $40\times40$ & \cmark / \xmark\\ 
		Al-antari {et al.} \cite{al2018fully} (2018) &$92.69$ & - & $40\times40$ &\cmark / \cmark \\ 
		{U-Net \cite{ronneberger2015u}}  (2015)  & $88.54\pm1.17$ & $83.85\pm2.13$ & $40\times40$ & \xmark / \xmark\\ 
		
		Li {et al.} \cite{li2018improved} (2018) & $\mathbf{93.66}\pm0.10$ & $\mathbf{92.23}\pm0.26$ & $40\times40$ & \xmark / \xmark\\ 
		\midrule
		
		{U-Net \cite{ronneberger2015u}}  (2015) & $89.79\pm0.33$ & $90.42\pm0.37$ & $224\times224$ & \xmark / \xmark\\ 
		Dhungel {et al.} \cite{dhungel2017deep} (2017) & $85$ & - & original & \cmark / \cmark\\ 
		
		Wang {et al.} \cite{wang2019multi} (2019) &$91.10$& {91.69} & $256\times256$ &\cmark / \cmark \\ 
		Singh {et al.} \cite{singh2020breast} (2020) & $92.11$ & - & $256\times256$ &\cmark / \cmark \\ 
		\midrule
		\dcn (2020)  & $\mathbf{93.69}\pm0.17$ & $\textbf{{92.17}}\pm0.03$ & $224\times224$ & \xmark / \xmark\\ 
		\bottomrule
	\end{tabular}
	\label{tab:segmentation_results2}
\end{table*}}

In particular, the fully connected CRF can be defined as follows:
\begin{equation}
\begin{aligned}
p(\mathbf{y}|h(\mathbf{x})) = \frac{1}{Z(h(\mathbf{x}))} \exp\bigg(\sum_{i\in V} \phi_u\big(h(\mathbf{x}_{i})\big) + \\ \sum_{i,j\in{E}}\phi_p\big(\mathbf{y}_{i},\mathbf{y}_j \mid h(\mathbf{x})\big)\bigg)
\end{aligned}
\label{eq:crunet_losscrf}
\end{equation}
where $Z$ is the partition function, and $h(\mathbf{x})$ is the deep latent feature of input $\mathbf{x}$ calculated by the softmax layer in CGL path. 
$\phi_u$ is the unary potential function, which is initialized as $h(\mathbf{x})$. 
$\phi_p$ is the pair-wise potential function which is formulated as  
\begin{equation}
    \phi_p\big(\mathbf{y}_i, \mathbf{y}_j \mid h(\boldsymbol{x})\big) = \mu\big(\mathbf{y}_i, \mathbf{y}_j\big)\sum_{m}\mathbf{w}^{(m)}k_G^{(m)}\big(\mathbf{x}_i, \mathbf{x}_j\big),
\end{equation}
where $\mathbf{y}_{i}$ and $\mathbf{y}_{j}$ are the predicted labels of connected nodes for position $i$ and $j$ respectively. 
$\mu(\mathbf{y}_i^{(n)},\mathbf{y}_j^{(n)})$ is the label compatibility defined by the Pott's Model \cite{potts1952some}.
$\mathbf{w}^{(m)}$ is the learned weight and $k_G^{(m)}$ is the pre-defined weighed Gaussians over feature vectors at position $i$ and $j$ \cite{krahenbuhl2011efficient, li2018improved}.

\begin{figure*}[!t]
	\normalsize
	\begin{center}
    	\includegraphics[width=0.85\linewidth]{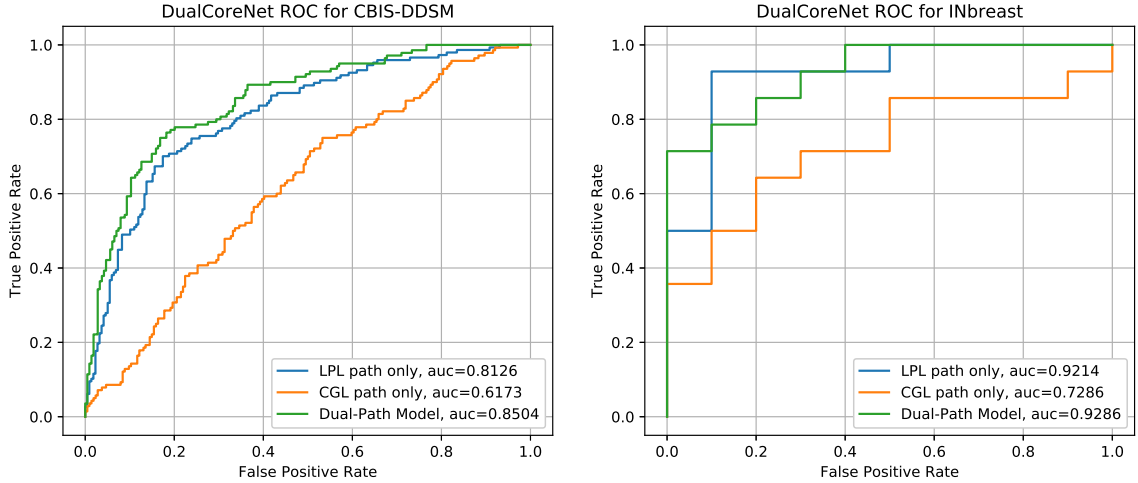}
		\caption[Classification ROC curves for INbreast and DDSM datasets by \dcnn]{Mass classification ROC curves of the \dcn experimented on INbreast and CBIS-DDSM for identifying the cancerous masses from the union of Benign and Malignant ROIs. }
	\label{fig:rocs_dcn}
	\end{center}
\end{figure*}

In order to improve the segmentation performance for the unbalanced mass contained ROIs, we proposed to minimize the below two dice losses $g$ and $f$ for the CGL path to output a high-resolutional binary label mask. 

\begin{equation}\label{eq:cgl}
\resizebox{.48\textwidth}{!}{${{\text{\large\ensuremath\ell}}}_{\text{CGL}} = \underbrace{1 - \displaystyle\frac{2\displaystyle\sum \mathbf{y} \cdot p(\mathbf{x}, \boldsymbol{\theta}_2)}{\displaystyle\sum\mathbf{y} + \displaystyle\sum p(\mathbf{x}, \boldsymbol{\theta}_2)}}_{\text{CNN}} + \gamma \cdot \underbrace{1 - \displaystyle\frac{2\displaystyle\sum \mathbf{y} \cdot  p(\mathbf{y}|h(\mathbf{x}))}{\displaystyle\sum\mathbf{y} + \displaystyle\sum  p(\mathbf{y}|h(\mathbf{x}))}}_{\text{CRF}},$}
\end{equation}

where  $p(\mathbf{x}, \boldsymbol{\theta}_2)$ is the output of the final softmax layer in the CGL path,  $\boldsymbol{\theta}_2$ are the parameters, $\gamma$ is the trade-off factor in the CGL path. 

\subsubsection{Fusion Module}
So far, the textual features and shape features have been extracted by the LPL and CGL path, it is natural to integrate these two separate features in the feature fusion block, in order to further improve the diagnosis performance. To do that, we propose a fusion module as shown in Fig. \ref{fig:dcn_architecture}. In particular, we first use two transformation blocks, each of which consists of several convolution layers followed by an average pooling and two fully connected layers, to transfer the output feature maps of LPL and CGL paths, respectively,
such transformed feature maps from two paths will feed into a  softmax layer to output the final classification result. The overall categorical cross-entropy based loss for classification task is defined as:
\begin{equation}\label{eq:fusion}
{{\text{\large\ensuremath\ell}}}_{\text{Fusion}} = -\sum_{n=1}^N\log p\big({z}^{(n)} \mid \boldsymbol{x}^{(n)};\boldsymbol{\theta} \big),
\end{equation}
where $z$ is the diagnosis class indicator and $\boldsymbol{\theta}$ is the entire network parameter vector.

Finally, by integrating the losses for LPL path \emph{e.g.} (\ref{eq:lpl}), CGL path \emph{e.g.} (\ref{eq:cgl}) and Fusion modules \emph{e.g.} {eq:fusion}, the entire \dcn loss is thereby defined as:
\begin{equation}\label{eq:dcn}
{{\text{\large\ensuremath\ell}}}_{\textsc{DualCoreNet}} = {{\text{\large\ensuremath\ell}}}_{\textrm{Fusion}} + \alpha \cdot {{\text{\large\ensuremath\ell}}}_{\textrm{LPL}} + \beta \cdot {{\text{\large\ensuremath\ell}}}_{\textrm{CGL}},
\end{equation}
where $\alpha$ and $\beta$ are  two trade-off factors to control the importance of $LPL$ and $CGL$ paths, which are empirically set as $\alpha = \beta = 1$ in our experiments.

\subsection{Implementation}
We use the Adam to optimize our \dcn. In order to alleviate overfitting and improve generalization, we use several training techniques for the \dcn model.   
(1) Regarding the initialization of the LPL path, the DDSM dataset is first trained in the LPL path and the parameters are then fine-tuned in the INbreast dataset. 
(2) The dropout layers are employed with 50\% random parameters dropping.

\section{Experiments}
\label{sec:4}
\subsection{Materials}
In this paper, we validate the proposed \dcn with two public mammography datasets:  CBIS-DDSM \cite{lee2017curated} and INbreast \cite{moreira2012inbreast}.
The CBIS-DDSM is a modernized subset of Digital Database for Screening Mammography (DDSM) \cite{heath2000digital}, in which 2478 digitized mammograms are formatted in DICOM format. 
On top of that, the CBIS-DDSM  has already been partitioned into the training (1318 masses) and test set (378 masses). In this paper, we adopt the same data division method as the website did. 
In the matter of class balance, either training or test set involves the equivalent amount of two classes of lesions.
Particularly, the malignant and benign ratio for the training and test sets are both roughly $1:1.07$.
The INbreast dataset \cite{moreira2012inbreast} is a Full Field Digital Mammography (FFDM) dataset, which was acquired at a Breast Centre in Hospital de São João, Porto, Portugal. There are a total of 115 cases in the INbreast dataset, which contains 410 mammogram images. Analogously, there are also two mammographic views for each breast and the images were annotated by human experts in a pixel-level labeling fashion. 
Regarding the data division for the evaluation of \dcnn, the INbreast data set is divided by patients into a training set and a test set as 80\%: 20\%.

In terms of the ROIs selection, masses are center cropped by two scales. 
One scale is the rectangular tight bounding box padded with 5 pixels on each boarder, which is utilized by the CGL to explore the segmentation related features.
The other scale is to crop the contextual rectangular region with proportional padding,  so that mass-centered ROI includes regions $2$ times the size of the bounding box. These contextual ROIs are utilized by the LPL path to extract latent and hierarchical features from masses and their surrounding tissues.
The selected ROIs are then all resized into identical dimension $224\times 224$ by bicubic interpolation. Accordingly, the ground truth binary masks are cropped and resized but with the nearest neighbor interpolation.
To avoid overfitting and provide better generalization, the selected ROIs are augmented with horizontal and vertical flips and random crop (with augmentation probability of 50\% for each instance) after data division.

\begin{figure}[!t]
    \centering
    \includegraphics[width=0.99\linewidth]{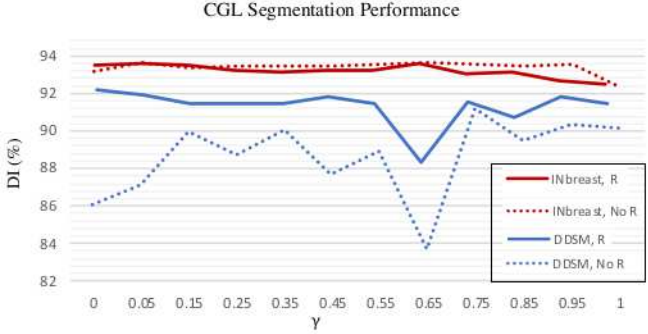}
\caption[CGL segmentation performance for DDSM and INbreast datasets in \dcnn]{CGL segmentation performance for DDSM and INbreast datasets in \dcnn. Dice coefficients (\%) are averaged over 10 experiments with different $\gamma$ in loss (\ref{eq:cgl}) and residual configurations ("R" represents residual and "No R" means without residual skips) on INbreast and DDSM datasets.}
\label{fig:cgl_dice_parameters}
\end{figure}

\subsection{Results and Analysis}
\subsubsection{Comparison with State-of-the-art}
We first compared the propose \dcn with five related state-of-the-art breast mass diagnosis methods \cite{arevalo2016representation, dhungel2016automated, kooi2017discriminating, dhungel2017deep, al-antari2020deep}. The results are listed in Table \ref{tab:dcn_all_comparison} where the performances of these compared methods are obtained from the results presented by their papers.
In particular, \cite{dhungel2017deep} and \cite{al-antari2020deep} have solved both breast mass segmentation and classification problem. 
Compared with these methods, \dcn is the only algorithm which has experimented on the DDSM dataset and \dcn has achieved leading diagnosis performance (second-best) on  INbreast dataset. 
\dcn produces a 0.93 and 0.85 AUC score for mass diagnosis on INbreast and DDSM dataset, respectively.
When compared with \cite{al-antari2020deep}, there is only a 0.001 AUC difference,
which is mainly because \cite{al-antari2020deep} randomly divided the training and test set after data augmentation.  In our paper, however, we first divide the original data into training and test set, which are followed by augmentation. 
Furthermore, \dcn has obtained the best segmentation performance when compared with all other algorithms, yielding 93.69\% and 92.17\% for INbreast and DDSM dataset.

Additionally, we evaluated the effectiveness of residual learning employed in \dcnn, we compare the test loss on two datasets between the configuration with or without residual skips in the best performing architecture (Fig. \ref{fig:dcn_architecture}). 
Generally, a desired validation loss is expected to be stable after some training epochs, after consistently decreasing with the increasing training epoch.
As shown in Fig. \ref{fig:validation_loss}, the validation loss of the vanilla \dcn either did not generalize or even slightly increased on both datasets.
On the contrary, the residual learning \dcn has shown a good ability of generalization, in which the validation loss was decreasing and saturated with the increasing number of training epochs. 
It is noted that the validation loss of the residual learning is much higher than that without residual skips (Figure \ref{fig:validation_loss}). This is caused by the weight decay regularization term. Since the number of parameters in the residual learning is larger than that in the no residual connection \dcn. 

In addition, as shown in Table \ref{tab:dcn_lpl_classification}, the cancer diagnosis performance of \dcn (Malignant vs Benign) with different regularization configurations (augmentation or pre-training) have been listed.
It can be noticed that the pre-training has markedly improved the model performance and augmentation method further competently  increased the generalization. 
When \dcn is trained with pre-training and data augmentation, the diagnosis AUC score for DDSM and INbreast has achieved 0.85 and 0.93, respectively.

\subsubsection{The importance of Dual-paths}
We are interested in studying the importance of dual-paths.  We evaluate the CGL and the LPL path for the  segmentation and classification performance, each of which is individually trained by ${{\text{\large\ensuremath\ell}}}_{\textrm{CGL}}$ and ${{\text{\large\ensuremath\ell}}}_{\textrm{LPL}}$ only.


\begin{figure}[!t]
    \centering
    \includegraphics[width=0.49\textwidth]{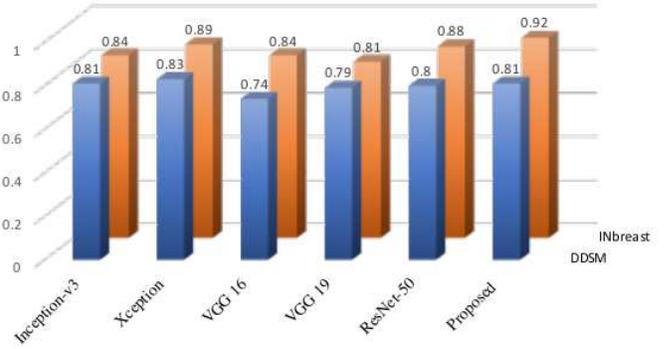}
    \caption[LPL diagnosis performance (Malignant vs Benign) by various backbone networks]{LPL diagnosis performance (Malignant vs Benign) by various backbone networks}
    \label{fig:lpl_backbones}
\end{figure}

Regarding the segmentation performance of CGL, the dice coefficients comparison between related works have been listed in Table \ref{tab:segmentation_results2}. For all the listed algorithms, only mass contained ROIs are provided as the input.
In terms of low-resolutional mass segmentation with output dimension $40 \times 40$, \cite{li2018improved} has achieved the best performance with 93.66\% and 92.23\% on INbreast and CBIS-DDSM, respectively.
With respect to the high-resolutional segmentation, the \dcn is so far the best performing algorithm for both datasets, with a 93.69\% DI score in INbreast and 92.17\% in DDSM dataset. 
Fig. \ref{fig:cgl_dice_parameters} has shown the segmentation performance when CGL is trained with various $\gamma$ values. It can be noticed that the over all segmentation performance on the INbreast dataset is significantly better than that on the DDSM dataset, which is mainly attributed to the higher quality data of INbreast. 
Specifically for the INbreast dataset, the no residual connections and residual  configuration generally performs equivalently. However, the best performance was obtained by no residual connection configuration when $\gamma$ is 0.65.  Note that the worst performance on INbreast (with either residual skips or not) was at $\gamma = 1$, where the CGL segmentation loss is contributed to the CNN and graphical model as ratio $1:1$. 
In terms of the DDSM dataset, the overall better performance was produced by the residual learning configuration. The best performance was obtained at $\gamma = 0.42$, where the segmentation DI achieves 92.17\%. 

The visualized segmentation results of CGL and radiologists' delineations can be seen in Fig. \ref{fig:ddsm_contours}  for both datasets.  
It can be noticed that, the proposed segmentation method performs very well with higher resolution mammograms, in which fine boundary details and irregular shape contours are both well depicted. There are no resulting spurious regions in \dcnn, which is mainly due to the structural consistency restriction by the graphical inference layer. 
Although we implement the graphical inference in a small spatial size before converting to high resolutions, the performance is not affected.
By doing so, a more efficient inference can be obtained with much less parameters and computing time.

As for the classification performance of the LPL path, we compared the LPL classification performance with various backbone networks, such as the Inception-v3, Xception, VGG16, VGG19, and ResNet-50 (Fig. \ref{fig:lpl_backbones}).
It can be seen that all networks perform better on the INbreast dataset. 
The best performances for INbreast and DDSM are obtained by the proposed LPL architecture (0.92 AUC score) and Xception network, respectively. However, the performance difference margin between our LPL (0.81 AUC score) and the Xception (0.83 AUC score) on DDSM is very small, with only a 0.02 AUC score. 
Generally speaking, the proposed LPL architecture achieved the leading performance for breast cancer diagnosis on both INbreast and DDSM datasets.

\subsubsection{Ablation study on the training loss}


Finally, We compared the \dcn with with different loss function configurations.:\footnote{Note we observed in our experiments that if we trained the model with only use ${{\text{\large\ensuremath\ell}}}_{\textrm{Fusion}}$ would lead to a not converged model.}    
${{\text{\large\ensuremath\ell}}}_{\textrm{Fusion}} + {{\text{\large\ensuremath\ell}}}_{\textrm{CGL}} $, 
${{\text{\large\ensuremath\ell}}}_{\textrm{Fusion}} + {{\text{\large\ensuremath\ell}}}_{\textrm{LPL}} $,
and  ${{\text{\large\ensuremath\ell}}}_{\textrm{Fusion}} + {{\text{\large\ensuremath\ell}}}_{\textrm{CGL}} + {{\text{\large\ensuremath\ell}}}_{\textrm{LPL}}$.
The diagnosis ROC curves of  \dcn with the Fusion loss combined with the LPL path, CGL path and dual-paths loss  are shown in Fig. \ref{fig:rocs_dcn}.
It can be noticed that \dcn performs best when activating both paths. 
And the second best performing training loss configuration is obtained by the Fusion and LPL path loss. 
This indicates that the features integrated from bot paths can learn richer information from the data.

\section{Conclusions}
\label{sec:5}
In this paper, we propose an innovative dual-path CNN architecture called \dcn for segmentation and classification problem. \dcn first embeds the original mammography data into two heterogeneous data domains (i.e. the original image and the binary mask domain), where deep features are  jointly learned. 
By integrating the conditional graph learner path and the locality preserving learner path,  our \dcn works in a simple but effective way to jointly learn segmentation and classification. 
The integrated intrinsic localized textural features and semantic information extracted from binary masks contribute to an interpretable and more discriminative representation, which can maximize the similarity margins between benign and malignant instances in the deep latent space.
Extensive experiments have shown that our method outperforms the state-of-the-arts on both breast mass segmentation and classification tasks in mammography. In addition \dcn performs better on higher quality dataset (the INbreast dataset). 

\bibliographystyle{IEEEtran}
\bibliography{refs}

\end{document}